\RequirePackage{fix-cm}
\documentclass[twocolumn]{svjour3}          
\smartqed  
\usepackage{graphicx}
\usepackage{mathptmx}       
\usepackage{helvet}         
\usepackage{courier}        
\usepackage{type1cm}        
\usepackage{makeidx}         
\usepackage{graphicx}        
\usepackage{multicol}        
\usepackage[bottom]{footmisc}

\usepackage{verbatim}

\usepackage{graphics} 
\usepackage{epsfig} 
\usepackage{mathptmx} 
\usepackage{times} 
\usepackage{amsmath} 
\usepackage{amssymb}  
\usepackage{float}

\begin{document}

\title{ Improvement of plant performance using Closed loop Reference Model Simple Adaptive Control  for Micro Air Vehicle
\thanks{The article was presented at the Euro GNC 2017- $4^{\text{th}}$ CEAS Specialist Conference on Guidance Navigation and Control 2017.}
}



\author{Shuvrangshu Jana         \and
        M. Seetharama Bhat 
}


\institute{Shuvrangshu Jana \at
              Indian Institute of Science, India \\
              Tel.: +91-8861551519\\             
              \email{shuvra.ce@gmail.com}           
           \and
            M. Seetharama Bhat \at
            Indian Institute of Science, India
}


\maketitle


\begin{abstract}

In this paper, we present a novel idea to improve the transient performance of the existing Simple Adaptive Control architecture, without requiring high adaptation gains. Improvement in performance is achieved by incorporating the closed loop reference model based on the output feedback to the Simple Adaptive Control architecture. In this proposed scheme, the reference model dynamics is driven by the desired command as well as the error signal between the plant output and the reference model output. It is shown that the modified control architecture improves the system performance without any additional control efforts, which is then validated through simulations of the lateral model dynamics of Micro Air Vehicle.  
\end{abstract}

\section{Introduction}

Micro Air Vehicle (MAV) dynamics is highly complex due to the coupling of longitudinal and lateral modes, propeller induced non-symmetric flow effects and motor counter torque effects, etc. The nominal velocity of MAVs is comparable to the wind velocity; hence MAVs are susceptible to wind gusts. MAV dynamics is highly uncertain due to uncertain system parameters and large disturbances from the environment; therefore, the control system of MAV should have high disturbance rejection property. If a single fixed controller is to guarantee the stability at all operating conditions, the performance is compromised. The adaptive controller adapts itself to the realistic situation by adjusting the appropriate control parameters \cite{lavretsky2013robust,ioannou2012robust,narendra2012stable}; thereby making it suitable for MAV control system design.

Simple Adaptive Control (SAC) is a direct adaptive control which is implemented using output feedback. Furthermore, SAC does not need persistent excitation and the reference model being of the same order as that of the plant \cite{kaufman2012direct,barkana2014simple,barkana2016robustness}. Since MAV states are not available accurately, therefore, SAC is a good candidate for MAV controller design.

In adaptive systems, the objective of asymptotic tracking can be achieved using the conventional Lyapunov method. However, the plant performance during the transient phase may not be satisfactory \cite{ioannou2012robust,narendra2012stable}. The transient performance can be improved by increasing the adaptation gain, but this leads to an increase in the frequency of the control signal; which may drive the overall system to instability. Further, this increase in the frequency of control input may not be desirable due to actuator bandwidth limitations and actuator fatigue. To improve the transient performance, the idea of modifying the reference model is considered by the introduction of the error feedback term to the reference model dynamics in standard Model Reference Adaptive Control (MRAC) structure\cite{lavretsky2013robust,stepanyan2010mrac,gibson2013closed,stepanyan2013output}. The reference model is made closed loop in nature by error feedback instead of traditional open loop reference model architecture. The main motivation is to drive the reference model towards the plant according to the error between the outputs of the plant and reference model. The driving of the reference model towards the plant reduces the magnitude and bandwidth of the control efforts required for the plant to follow the reference model.

The novelty of this paper is in extending the concept of closed loop reference model to existing SAC architecture by introducing output error feedback term to the reference model dynamics. We denote the new architecture as Closed loop reference model Simple Adaptive Control (CL-SAC)\cite{jana2017closed}. It is found that with a proper selection of output error feedback gain, CL-SAC structure improves the transient performance without any additional control efforts. The new output error feedback term increases the damping in reference model dynamics, thus making it faster to learn the plant parameters. The improvement in performance is validated through the simulation of MAV lateral model dynamics.

The rest of the paper is organized as follows: Section 2 describes the existing SAC architecture. 
The new CL-SAC structure is described in section 3. Performance improvement of CL-SAC is shown through analysis in section 4. Simulation of lateral dynamics of MAV is performed in section 5. Section 6 summarizes the result.

\section{ SAC Architecture}
\begin{figure}[H]
\begin{flushleft}
\includegraphics[width=1.0\linewidth, height= 1.0\linewidth]{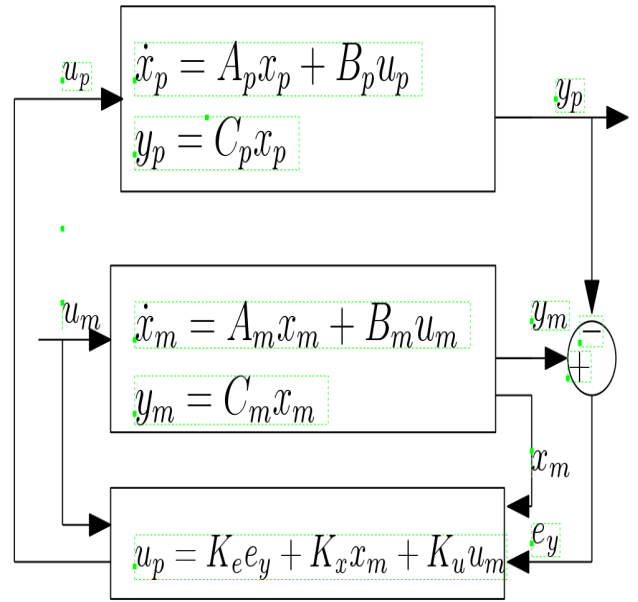}    
\caption{Block diagram of basic SAC architecture} 
\label{fig:fig1}
\end{flushleft}
\end{figure}

SAC is a special adaptive control methodology based on model following concept, where a higher dimensional plant can track the output of a lower order reference model with a lower order controller. It has been successfully applied in missiles \cite{barkana2005classical}, spacecraft \cite{maganti2007simplified}, boost converters \cite{jeong2011design}, space manipulators \cite {ulrich2014nonlinear}, quadcopter \cite{tomashevich2017simple} etc. Its reference model represents the input-output behaviour of the plant, and it is sufficient to generate the desired command to be tracked. This methodology requires that the plant is ``W-almost strictly passive" and the plant transfer function is ``W-almost strictly positive real (WASPR)"; therefore it is stabilizable through some positive definite output feedback \cite{barkana2016adaptive}. SAC architecture assumes that the primary stability properties of the plant are available. The basic stability information about the plant can be used to make the plant WASPR with parallel feedforward configuration\cite{rusnak2012duality}. Consider the plant to be linear time-invariant m x m square systems as
\begin{equation}
  \dot{x}_{p}=A_{p}x_{p}+B_{p}u_{p}  \label{plant}
\end{equation}
\begin{equation}
  y_{p} =C_{p}x_{p}
\end{equation}
where $ x_{p} \in \mathbb{R}^{n_{p}}$, $u_{p} \in \mathbb{R}^{m}$, $y_{p} \in \mathbb{R}^{m}$ are the plant states, inputs and outputs respectively.\\
The control objective is to ensure that the  plant output tracks the commanded bounded reference signal while tracking the output of the specified reference model. The reference model is designed as per system response specification and is driven by reference signal to generate  the desired command  for the plant. 
In basic SAC architecture, the following reference model is considered 
\begin{equation}
  \dot{x}_{m}=A_{m}x_{m}+B_{m}u_{m} \label{olrm1}
\end{equation}
\begin{equation}
  y_{m} =C_{m}x_{m}\label{olrm2}
\end{equation}
where $ x_{m} \in \mathbb{R}^{n_{m}}$, $u_{m} \in \mathbb{R}^{m_{m}}$, $y_{m} \in \mathbb{R}^{m}$  are  the reference model states, reference signals and reference model outputs respectively. $A_{m}$ is chosen as a Hurwitz matrix and  the reference model assumed to be bounded input bounded output (BIBO) stable.\\
     The reference model dimension $ n_{m} $ can be less than  plant dimension $ n_{p}$. However, the output dimension of both plant and reference model should be the same, as the plant output has to track the reference model output. The output tracking error is defined as  
\begin{equation}
  e_{y} =y_{m}-y_{p}.
\end{equation}
\noindent The control law is defined  based on the command generator tracker concept \cite{broussard1980feedforward} as 
\begin{equation}
  u_{p} =K(t)r \label{saccontrol}
\end{equation}
 \noindent where $ K(t) =[K_{e}(t), K_{x}(t), K_{u}] $ and 
 \begin{equation*}
       r= \left[ \begin {array}{c} e_{{y}}  
\\ \noalign{\medskip}x_{{m}}  \\ \noalign{\medskip}u_{
{m}}  \end {array} \right]
\end{equation*}
\begin{equation*}
 K_{e}(t)= K_{pe}(t)+K_{Ie}(t) 
\end{equation*}
\begin{equation*}
  K_{pe}(t)= e_{y}e_{y}^{T}\Gamma_{pe}; \quad
 \dot K_{Ie}(t)= e_{y}e_{y}^{T}\Gamma_{Ie} -\sigma K_{Ie}
\end{equation*}
\begin{equation*}
 K_{x}(t)= K_{px}(t)+K_{Ix}(t)
\end{equation*}
\begin{equation*}
  K_{px}(t)= e_{y}x_{m}^{T}\Gamma_{px}; \quad
 \dot K_{Ix}(t)= e_{y}x_{m}^{T}\Gamma_{Ix}
\end{equation*}
\begin{equation*}
 K_{u}(t)= K_{pu}(t)+K_{Iu}(t)
\end{equation*}
\begin{equation*}
  K_{pu}(t)= e_{y}u_{m}^{T}\Gamma_{pu}; \quad
 \dot K_{Iu}(t)= e_{y}u_{m}^{T}\Gamma_{Iu}
\end{equation*}
\noindent $ \Gamma_{pe}$ , $\Gamma_{Ie}$ , $\Gamma_{px}$ , $\Gamma_{Ix}$ , $\Gamma_{pu}$ , $\Gamma_{Iu}$ are the time invariant weighing matrices of appropriate dimension, $K_{Ie}(0)=K_{Ie0}$ , $K_{Ix}(0)=K_{Ix0}$ , $K_{Iu}(0)=K_{Iu0}$. $\sigma $ is a positive scalar adjusted such that it does not allow the integral gain to increase without bound.
\noindent The block diagram of SAC architecture is given in Fig.\ref{fig:fig1}.  
      When system output tracks the reference model output perfectly, plant state and control trajectories follow the ideal state and control trajectories, denoted by  $ x_{p}^{*} $ and $u_{p}^{*}$. The ideal state and control trajectories are 
\begin{equation}
  \dot{x}_{p}^{*}=A_{p}x_{p}^{*}+B_{p}u_{p}^{*} \label{ideal}
\end{equation}
\begin{equation}
 u_{p}^{*} =  \tilde{K_{x}}x_{m}+ \tilde{K_{u}}u_{m}. 
\end{equation}
where $ \tilde{K_{x}}$ and $ \tilde{K_{u}}$ are unknown ideal control gains.
In perfect tracking case, 
\begin{equation}
 y_{p}^{*}=y_{m}=C_{p}x_{p}^{*}=C_{m}x_{m}. \label{perfecttracking}
\end{equation} 
\noindent The error between ideal state and plant state is
\begin{equation}
  e_{x}=x_{p}^{*}-x_{p}.
\end{equation}
Similarly the output error is
\begin{equation}
 e_{y}=C_{p}x_{p}^{*}-C_{p}x_{p}=C_{p}e_{x}.
\end{equation} 
\noindent After deducting (\ref{plant}) from (\ref{ideal}), the error differential equation is obtained as  
 \begin{equation}
 \dot{e_{x}} =(A_{p}-B_{p} \tilde{K}_{e}C_{p})e_{x}-B_{p}(K(t)-\tilde{K})r \label{errord}
\end{equation} 
where $ \tilde K(t) =[\tilde K_{e}, \tilde K_{x},\tilde K_{u}] $.
The following  Lyapunov function is considered for stability of error dynamics.
\begin{equation}
 V(t)=e_{x}^{T}Pe_{x}+trace[W((K(t)-\tilde{K})\Gamma^{-1}(K(t)-\tilde{K})^{T}] \label{Lyapunov}
\end{equation} 
where P and Q  are positive definite symmetric matrices, W is a positive definite matrix and these matrices are the solution of the following two equations.
\begin{equation}
 P(A_{p}-B_{p} \tilde{K}_{e}C_{p})+(A_{p}-B_{p} \tilde{K}_{e}C_{p})^{T}P=-Q \label{wasprc1}
\end{equation} 
\begin{equation}
PB_{p}=C_{p}^{T}W^{T}. \label{wasprc2}
\end{equation} 
 The conditions (\ref{wasprc1}) and (\ref{wasprc2}) are  known as W-ASPR conditions. If the  plant is  minimum phase and the eigenvalues of its input output matrix product ($ C_{p}B_{p} $) are located in the right half plane, then it satisfies the W-ASPR conditions. Real world plants are not necessarily W-ASPR, but they can be made W-ASPR after augmenting them with parallel feedforward configurations (PFC). The augmented plant $P(s) + D(s)$ is W-ASPR if it is relative degree of zero or one, where $ D(s)=  \frac{1}{C(s)}$ and  $C(s) $ stabilizes the plant P(s) through feedback configuration.
\noindent In the case of W-ASPR plant, the derivative of candidate Lyapunov equation (\ref{Lyapunov}) along the trajectory of error dynamics (\ref{errord}) reduces to
\begin{equation}
\dot{V}(t)=-e_{x}^{T}Qe_{x}\leq 0.
\end{equation}

As the derivative of the candidate Lyapunov function is non-positive,the error dynamics is stable. In this case, $V(t)$ is lower bounded, $\dot{V}(t)$ is negative semi definite and $\dot{V}(t)$ is uniformly continuous in time. There for using ``Lyapunov-like lemma" \cite{slotine1991applied}, it can be concluded that $\displaystyle{\lim_{ t\to \infty}} \dot{V}(t)=0$. Hence the error between the plant and the ideal trajectory vanishes asymptotically. Therefore, 
$C_{p}x_{p}=C_{p}x_{p}^{*}=C_{m}x_{m} $. So, the plant output follows the reference model output

\section{CL-SAC architecture }
The closed loop reference model gives better performance than the open loop part in MRAC architecture \cite{lavretsky2013robust,gibson2013adaptive,wiese2015adaptive}. Closed loop reference model structure is based on the concept that driving the reference model towards the plant will reduce the control efforts required for the plant to track the reference model. The similar modification in the reference model is made in the basic SAC architecture to improve the transient performance of the original structure.  In this paper, the modification of the reference model is based on the output error and it is applicable to general MIMO square systems. In CL-SAC, the reference model is not only driven by the reference signal but also by the output error between the plant and the reference model. In this case, the open loop reference model is
\begin{equation}
  \dot{x}_{m}=A_{m}x_{m}+B_{m}u_{m}.  \label{olrm}
\end{equation}
\noindent The modified closed loop reference model is considered as follows 
\begin{equation}
  \dot{x}_{mo}=A_{m}x_{mo}+B_{m}u_{m} - L_{v}(y_{mo}-y_{p}) \label{clrm1}
\end{equation}
\begin{equation}
  y_{mo} =C_{m}x_{mo} \label{clrm2}
\end{equation}
 The value of $ L_{v}$ is chosen such that the  reference model and subsequent error dynamics are BIBO stable. This reference model is termed as closed loop reference model and earlier reference model in equation (\ref{olrm1})-(\ref{olrm2}) is denoted as open loop reference model.  In this case output tracking error is
\begin{equation}
  e_{my} =y_{mo}-y_{p}.
\end{equation}
The controller structure is defined similar to the  basic SAC architecture, where reference  model states and outputs are calculated from the modified reference model dynamics (\ref{clrm1}) and (\ref{clrm2}).
\begin{equation}
  u_{p} =K(t)r
\end{equation}
\noindent  where,
\begin{eqnarray*}
 K(t) &=& [K_{e}(t), K_{x}(t), K_{u}(t)] \\
 r &=&  [e_{my}, x_{mo}, u_{m}]' \\
 K_{e}(t)&=& K_{pe}(t)+K_{Ie}(t); \\
 K_{pe}(t)&=& e_{my}e_{my}^{T}\Gamma_{pe}; \quad
 \dot K_{Ie}(t)= e_{my}e_{my}^{T}\Gamma_{Ie} -\sigma K_{Ie} \\
 K_{x}(t)&=& K_{px}(t)+K_{Ix}(t); \\ 
 K_{px}(t)&=& e_{my}x_{mo}^{T}\Gamma_{px}; \quad
 \dot K_{Ix}(t)= e_{my}x_{mo}^{T}\Gamma_{Ix};   \\
 K_{u}(t)&=& K_{pu}(t)+K_{Iu}(t); \\
 K_{pu}(t)&=&  e_{my}u_{m}^{T} \Gamma_{pu}; \quad
 \dot K_{Iu}(t)= e_{my}u_{m}^{T}\Gamma_{Iu}.
\end{eqnarray*}

\begin{figure}[H]
\begin{flushleft}
\includegraphics[width=1.0\linewidth, height= 1.0\linewidth]{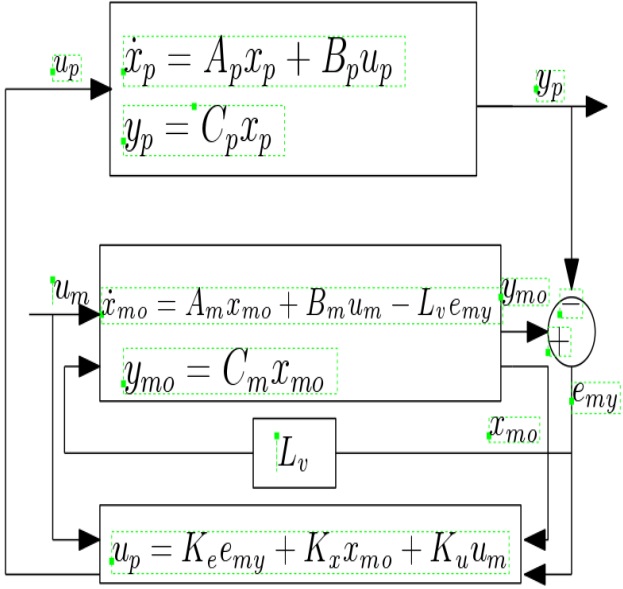}    
\caption{Block diagram of CL-SAC architecture} 
\label{fig:fig2}
\end{flushleft}
\end{figure}

 The initial conditions are kept the same as that of SAC architecture.
The basic  block diagram of CL-SAC architecture is given in  Fig. \ref{fig:fig2} . 
\subsection{Output model following }
In this section, the condition for output model following  for constant step input is revisited as the reference model dynamics is modified. The ideal trajectories are the linear functions of model states and model inputs.
\begin{equation}
  {x}_{p}^{*}=S_{11}x_{mo}+S_{12}u_{m} \label{modelfollow1}
\end{equation}
\begin{equation}
 u_{p}^{*} =  S_{21}x_{mo}+  S_{22}u_{m}. \label{modelfollow2}
\end{equation}
 From (\ref{ideal}), (\ref{perfecttracking}), (\ref{modelfollow1}) and  (\ref{modelfollow2}) ideal trajectory dynamics can be written as,
\begin{equation}
  \left[ \begin {array}{cccc} \dot{x}_{p}^{*}
\\ \noalign{\medskip} {y}_{p}^{*} \\ 
\end {array} \right] = \left[ \begin {array}{cccc} A_{p}& B_{p}
\\ \noalign{\medskip}C_{p} & 0\\ \noalign{\medskip}
\end {array} \right]  \left[ \begin {array}{cccc} S_{11}& S_{12}
\\ \noalign{\medskip}S_{21} & S_{22}\\ \noalign{\medskip}
\end {array} \right] \left[ \begin {array}{cccc} {x}_{m0}
\\ \noalign{\medskip} {u}_{m} \\ 
\end {array} \right] \label{idealmat}
\end{equation}
Let the reference command $u_{m}$ be restricted for a constant step input. After differentiating (\ref{modelfollow1}) gives
 \begin{equation}
 \dot{x}_{p}^{*}=S_{11}\dot x_{mo} \label{xpder}
\end{equation}
as $u_{m}$ is considered as a constant input.
Substituting the value of  $\dot x_{mo}$ from  (\ref{clrm1}) in (\ref{xpder}) gives
 \begin{equation}
 \dot{x}_{p}^{*}=S_{11}(A_{m}x_{mo}+B_{m}u_{m} - L_{v}(y_{mo}-y_{p})) \label{xpder2}
\end{equation}
In perfect tracking case, $ y_{p}$ is denoted as $ {y}_{p}^{*}$, then (\ref{xpder2}) can be simplified as
 \begin{eqnarray}
 \dot{x}_{p}^{*}=(S_{11}A_{m}-L_{v}C_{m})x_{mo}+S_{11}B_{m}u_{m} +S_{11}L_{v}C_{p}{x}_{p}^{*}\label{xpder3}
\end{eqnarray} 
Substituting the value of ${x}_{p}^{*}$ from (\ref{modelfollow1}) in (\ref{xpder3}) gives
\begin{eqnarray}
 \dot{x}_{p}^{*}=(S_{11}A_{m}-L_{v}C_{m}+S_{11}L_{v}C_{p}S_{11})x_{mo}+(S_{11}B_{m} \\\nonumber
 +S_{11}L_{v}C_{p}S_{12})u_{m}  \label{xpder4}
\end{eqnarray} 
In case of perfect tracking,
\begin{equation}
 y_{p}^{*}=y_{mo}=C_{p}x_{p}^{*}=C_{m}x_{mo}. \label{pccondition}
\end{equation} 
 Equation(\ref{xpder4}) and (\ref{pccondition}) can be combined as
\begin{equation} 
  \left[ \begin {array}{cccc} \dot{x}_{p}^{*}
\\ \noalign{\medskip} {y}_{p}^{*} \\ 
\end {array} \right]=\left[ \begin {array}{cccc} S_{11}A_{m}-L_{v}C_{m}+S_{11}L_{v}C_{p}S_{11}& S_{11}B_{m}+S_{11}L_{v}C_{p}S_{12}
\\ \noalign{\medskip}C_{m} & 0\\ \noalign{\medskip}
\end {array} \right]    \label{idealmatmod} 
\end{equation}
From  (\ref{idealmat}) and (\ref{idealmatmod}), (\ref{idealmatmod}) can be written as\\
\begin{eqnarray*}
\left[ \begin {array}{cccc} S_{11}A_{m}-L_{v}C_{m}+S_{11}L_{v}C_{p}S_{11}& S_{11}B_{m}+S_{11}L_{v}C_{p}S_{12}
\\ \noalign{\medskip}C_{m} & 0\\ \noalign{\medskip}
\end {array} \right] \\= 
\left[ \begin {array}{cccc} A_{p}& B_{p}
\\ \noalign{\medskip}C_{p} & 0\\ \noalign{\medskip}
\end {array} \right]  \left[ \begin {array}{cccc} S_{11}& S_{12}
\\ \noalign{\medskip}S_{21} & S_{22}\\ \noalign{\medskip}
\end {array} \right]
\end{eqnarray*}
 The solution of the above matrix equation exist as there are $ (n_{p}+m)(n_{m}+m_{m})$ number of equations as well as unknowns. This implies that perfect tracking in the case of constant step input is possible in case of modified architecture.\\
  If the reference command is generated from the following differential equation,
 \begin{equation}
\dot v_{m}=A_{v}v_{m} ; \quad 
u_{m}=c_{v}v_{m}
\end{equation}
 the perfect tracking is expected for following conditions
\begin{equation}
 dim(x_{m})+dim(u_{m})\geq dim(v_{m}).
 \end{equation}
 In case of general tracking command, it can be shown that CL-SAC architecture follows the reference command with bounded tracking error. 
\subsection{Stability analysis}
In this section, we analyze the stability of the system since the reference model dynamics is modified.This modification in the reference model does not affect the error dynamics (\ref{errord}).
\begin{equation}
  e_{x}=x_{p}^{*}-x_{p}
\end{equation}
\begin{equation}
 \dot{e_{x}} =(A_{p}-B_{p} \tilde{K}_{e}C_{p})e_{x}-B_{p}(K(t)-\tilde{K})r
\end{equation} 
Hence, it can be shown that the error between plant and ideal state  trajectory goes down asymptotically. Therefore, the output error between plant and closed loop reference model goes to zero, and hence, the closed loop model reduces to open loop reference model asymptotically. So, the plant output is able to track the reference model output.

The effect of the modified architecture is visible in output error dynamics between the plant output and the reference model output. This is discussed in details in the next section.

\section{ Performance analysis}
In the case of SAC architecture, the error dynamics between the plant output and the reference model output is obtained as follows:
\begin{equation}
\dot{e_{y}}= C_{m}\dot{x}_{m}-C_{p}\dot{x}_{p}. \label{outputerrordersac}
\end{equation}
\noindent Using the value of $\dot{x}_{m}$ and $\dot{x}_{p}$ from  (\ref{olrm1}) and (\ref{plant}) respectively in (\ref{outputerrordersac}); we get
\begin{equation}
  \dot{e_{y}}
=C_{m}(A_{m}x_{m}+B_{m}u_{m})-C_{p}(A_{p}x_{p}+B_{p}u_{p}) \label{outputerrordersac2}
\end{equation}
\noindent Putting the value of $u_{p}$ as per (\ref{saccontrol}) in (\ref{outputerrordersac2}), we get
\begin{eqnarray}
\dot{e_{y}}= 
        C_{m}(A_{m}x_{m}+B_{m}u_{m})-C_{p}(A_{p}x+B_{p}(K_{e}(t)e_{y} \\+
                     K_{x}(t)x_{m}+K_{u}u_{m})
\end{eqnarray}
\noindent Finally, we get
\begin{equation}
\dot{e_{y}}=A_{mm}e_{y}+\tilde{\theta}^{T}(t) \omega  \label{outputerrordersac3}
\end{equation} 
where $ A_{mm}=-C_{p}B_{p}K_{e}$, \\ 
\noindent $ \tilde{\theta}^{T}(t) = [-C_{p}A_{p} ,C_{m}A_{m}-C_{p}B_{p}K_{x}(t),C_{m}B_{m}- C_{p}B_{p}K_{u}(t)]$  and $\omega =[x_{p} ,x_{m} , u_{m} ]^{T}$. \\

In the case of CL-SAC architecture, the output tracking error is $e_{my}$ and  the error dynamics is
\begin{equation}
  \dot{e_{my}}
=C_{m}(A_{m}x_{mo}+B_{m}u_{m}-L_{v}(y_{mo}-y_{p}))-C_{p}(A_{p}x_{p}+B_{p}u_{p}).\label{outputerrorderclsac}
\end{equation}
\noindent Equation (\ref{outputerrorderclsac}) can be written as:
\begin{equation}
\dot{e_{my}}=A_{mn}e_{my}+\tilde{\theta}^{T}(t) \omega
\end{equation} 
where $ A_{mn}=-C_{p}B_{p}K_{e}-C_{m}L_{v}$, \\
   
\noindent $ \tilde{\theta}^{T}(t) = [-C_{p}A_{p}, C_{m}A_{m}-C_{p}B_{p}K_{x}(t), C_{m}B_{m}- C_{p}B_{p}K_{u}(t)]$   and $\omega =[x_{p} ,x_{mo} , u_{m} ]^{T}$. \\

\noindent  The plant is able to do bounded tracking of the output of the reference model, so the output error dynamics is stable.\\
In the case of SAC architecture, since the output error dynamics is stable; there exist a Lyapunov function V(t) whose derivative along the trajectories of the system in  (\ref{outputerrordersac3}) is negative semi-definite:
\begin{equation}
\dot V(t)=- e_{y}^{T}Q_{1}e_{y} \leq0
\end{equation}
where $Q_{1}=Q_{1}^{T}>0$. Let's consider this  Lyapunov function  in following form
\begin{equation}
V(t)=e_{y}^{T}P_{1}e_{y}+f(\tilde{\theta}(t),\Gamma)
\end{equation}
 where $ \Gamma $ represent the set of gains $ \Gamma_{Ix}, \Gamma_{Ie}$ etc and  $P_{1}=P_{1}^{T}>0$ is the solution of the following Algebraic Riccati equation.
\begin{equation}
 P_{1}A_{mm}+A_{mm}^{T}P_{1}=-Q_{1}
\end{equation} 
Since, $\dot{V}(t)\leq 0$;
\begin{equation}
\parallel e_{y} \parallel^{2}\leq \frac{V(t)}{\lambda_{min}(P_{1})}\leq \frac{V(0)}{\lambda_{min}(P_{1})} \label{liainequality}
\end{equation} 
where $\lambda_{min}(P_{1})$ is minimum eigenvalue value of $ P_{1} $.
 Since the error dynamics is bounded, $ f(\tilde{\theta}(t))$ is bounded by positive constant $ C $. So, bound for V(t) can be written as;
 \begin{equation}
V(t)\leq e_{y}^{T}P_{1}e_{y}+C \label{l1}
\end{equation}
\noindent  We have the following inequality,\\
  $ e_{y}^{T}Q_{1}e_{y} \geq e_{y}^{T} \lambda_{min}(Q_{1})e_{y} \geq \frac{\lambda_{min}(Q_{1})}{\lambda_{max}(P_{1})}e_{y}^{T}\lambda_{max}(P_{1})e_{y} $ \\
  
\noindent Further,\\
$ \frac{\lambda_{min}(Q_{1})}{\lambda_{max}(P_{1})}e_{y}^{T}\lambda_{max}(P_{1})e_{y} \geq \frac{\lambda_{min}(Q_{1})}{\lambda_{max}(P_{1})} e_{y}^{T}P_{1}e_{y}=\gamma (e_{y}^{T} P_{1}e_{y})$ \\
\noindent  where $\gamma =\frac {\lambda_{min}(Q_{1})} {\lambda_{max}(P_{1})}$. Therefore,
 \begin{equation}
\dot V(t)=- e_{y}^{T}Q_{1}e_{y} \leq -\gamma (e_{y}^{T}P_{1}e_{y}). \label{l2}
\end{equation}
Using the value of $e_{y}^{T}P_{1}e_{y}$ from (\ref{l1}) in (\ref{l2}) gives
\begin{equation}
\dot V(t)+\gamma V(t) \leq \gamma C.
\end{equation}   
\noindent After integration of $\dot V(t)$ we get,
\begin{equation}
 V(t)\leq V(0)e^{-\gamma t}+C -C e^{-\gamma t}.
\end{equation}
\noindent From (\ref{l1}), it follows that
\begin{equation}
V(0)-C \leq e_{y}^{T}(0)P_{1}e_{y}(0).
\end{equation}
\noindent Therefore $V(t)$ is upper bounded as
\begin{equation}
V(t) \leq C+ e_{y}^{T}(0)P_{1}e_{y}(0)e^{-\gamma t}.
\end{equation}
Considering (\ref{liainequality}), the output error has the following upper bound:
\begin{equation}
\parallel e_{y} \parallel \leq \sqrt{ (\lambda_{min}(P_{1}))^{-1} (C+ e_{y}^{T}(0)P_{1}e_{y}(0)e^{-\gamma t}) }
\end{equation}
One part of the above bound is due to initialization error between the plant output and the reference model output,  this term decays exponentially. If initialization is done perfectly, this output error bound reduces to,
\begin{equation}
\parallel e_{y} \parallel\leq \sqrt{C (\lambda_{min}(P_{1}))^{-1}}.
\end{equation}
\noindent $ \lambda_{min}(P_{1}) $ satisfies the following inequality \cite{yasuda1979upper}:  
\begin{equation}
 \lambda_{min}(P_{1}) \geq  2\lambda_{max}^{-1}(S_{1})
\end{equation}  
where  $S_{1} =-(A_{mm}+A_{mm}^{T})Q_{1}^{-1}/2$ and $ \lambda_{max}(S_{1})$ is the maximum eigenvalue of $S_{1}$.

\noindent Therefore, the output error satisfy the following bound:
\begin{equation}
\parallel e_{y} \parallel\leq \beta \sqrt{\lambda_{max}(S_{1})}
\end{equation}
where $\beta$ is constant depends on value of C.
\noindent The similar bound for the CL-SAC architecture is 
 \begin{equation}
\parallel e_{my} \parallel\leq \beta \sqrt{\lambda_{max}(S_{2})}
\end{equation}
where $S_{2} =-(A_{mn}+A_{mn}^{T})Q_{2}^{-1}/2$.\\
\noindent Let $Q_{1}=Q_{2} $ is  chosen as Identity matrix.\\ 
\noindent $\lambda_{max}(S_{2})= \lambda_{max}(-[(A_{mm}+A_{mm}^{T})-C_{m}L_{v} -(C_{m}L_{v})^{T} ])$\\
Clearly with proper choice of $L_{v}$,   $\lambda_{max}(S_{2}$)  can be made smaller than the $\lambda_{max}(S_{1}$). Therefore, the bound of $e_{my}$  can be made tighter than the $e_{y}$. Hence, in CL-SAC architecture output tracking error can be made further smaller than SAC architecture, which will improve the transient performance.
\subsection{ Selection of $ L_{v}$}
 In \cite{gibson2013adaptive}, some guidelines are available for design of the gain of the error feedback term for the case of MRAC structure. In the case of SAC structure similar analysis is performed. A  simplified analysis is done  for scalar reference model, i.e  the dimension of  $n_{m}, m_{m}, m $ is one. The response of the open loop reference model in (\ref{olrm}) with initial conditions $x_{m}(0)= x(0)$ is
\begin{equation}
x_{m}(t)=e^{A_{m}t}x(0)+ \int_0^t e^{A_{m}(t-\tau)} B_{m} u_{m} (\tau) d\tau
\end{equation}

\noindent The response of the closed loop reference model in (\ref{clrm1}) with same initial condition as  $x_{mo}(0)= x(0)$ is

\begin{eqnarray}
x_{mo}(t)=e^{A_{m}t}x(0)+ \int_0^t e^{A_{m}(t-\tau)} B_{m} u_{m} (\tau) d\tau \\ \nonumber 
-L_{v} \int_0^t e^{A_{m}(t-\tau)}e_{my}(\tau) d\tau
\end{eqnarray}

\noindent The difference of response between closed loop reference model in (\ref{clrm1})  and the open loop  reference model in (\ref{olrm}) having same initial conditions can be written as
\begin{equation}
\bigtriangleup x_{mo}(t) = x_{mo}(t)-x_{m}(t)=-L_{v} \int_0^t e^{A_{m}(t-\tau)}e_{my}(\tau) d\tau
\end{equation}

 \noindent Applying  Cauchy Schwartz inequality;
\begin{equation}
\int_0^t e^{A_{m}(t-\tau)}e_{my}(\tau) d\tau \leq  \sqrt{\int_0^t ( e^{A_{m}(t-\tau)})^2 d\tau} \sqrt{\int_0^t ( e_{my}(\tau))^2 d\tau} \label{ineq0}
\end{equation}

\noindent  We have,
\begin{equation}
 \int_0^t ( e^{A_{m}(t-\tau)})^2 d\tau \leq  \frac{1}{\parallel 2 A_{m} \parallel}\end{equation}   \label{ineq1}
Also,
\begin{equation}
\int_0^t ( e_{my}(\tau))^2 d\tau  \leq \frac{V(0)}{\lambda_{min}(P_{1})} \label{ineq2}
\end{equation} 

Using the inequalities in (\ref{ineq1}) and (\ref{ineq2}), the inequality in (\ref{ineq0}) can be simplified as,
\begin{equation}
\bigtriangleup x_{mo}(t) \leq \parallel L_{v}\parallel \sqrt{\frac{1}{\parallel 2A_{m}\parallel}}\sqrt{\frac{V(0)}{\lambda_{min}(P_{1})}} 
\end{equation}
\noindent Equivalently,
\begin{equation}
\bigtriangleup x_{mo}(t) \leq \parallel L_{v}\parallel \sqrt{\frac{1}{\parallel 2A_{m}\parallel}}\sqrt{\frac{e_{y}(0)^{T}P_{1}e_{y}(0)+f(\tilde{\theta}(0),\Gamma)}{\lambda_{min}(P_{1})}} \label{eqnf}
\end{equation}

The value of $L_{v}$ is chosen such that the reference model dynamics and subsequent error dynamics is BIBO stable.  Clearly from the bound of $\bigtriangleup x_{mo}(t)$; the high value of $L_{v} $  will make the $\bigtriangleup x_{mo}(t)$ large, i.e, the response of closed loop reference model will deviate more from the response of open loop reference model. Also, The low value of $L_{v}$ will reduce the effect of error feedback in the reference model dynamics.   The value of  $L_{v}$ needs to be selected as a function of  $\Gamma$ optimally to balance both the effects.  Mathematically, it can be shown that a high value of $L_{v} $ will reduce the oscillation in the control signal. In general, if the magnitude of $L_{v} $  is selected as high as possible and closer to the value of $\Gamma $, the advantage of $L_{v} $ can be obtained.

\section{Simulation}
MAV lateral dynamics are considered to verify the improvement in the new architecture. MAV lateral  states are lateral velocity (v),  roll rate (p), yaw rate (r), roll attitude $(\phi)$. Generally, in MAV case, roll rate and yaw rate are available form the gyroscope sensor and the $(\phi)$ is estimated from the estimation loop. The lateral dynamics need to track the roll attitude command from the guidance loop. Therefore, the plant output $y_{p}$ $(\phi)$ has to track the output of the reference model which is driven by the commanded roll angle ($\phi_{cmd})$. The roll rate and yaw rate can be feedback to increase the damping in the inner loop. Here, we will consider the tracking and stabilization problem only using the attitude $(\phi)$.  
\subsection{Lateral dynamics-Roll attitude tracking}
  
\begin{figure}[H]
\begin{flushright}
\includegraphics[width=1.0\linewidth, height= 0.7\linewidth]{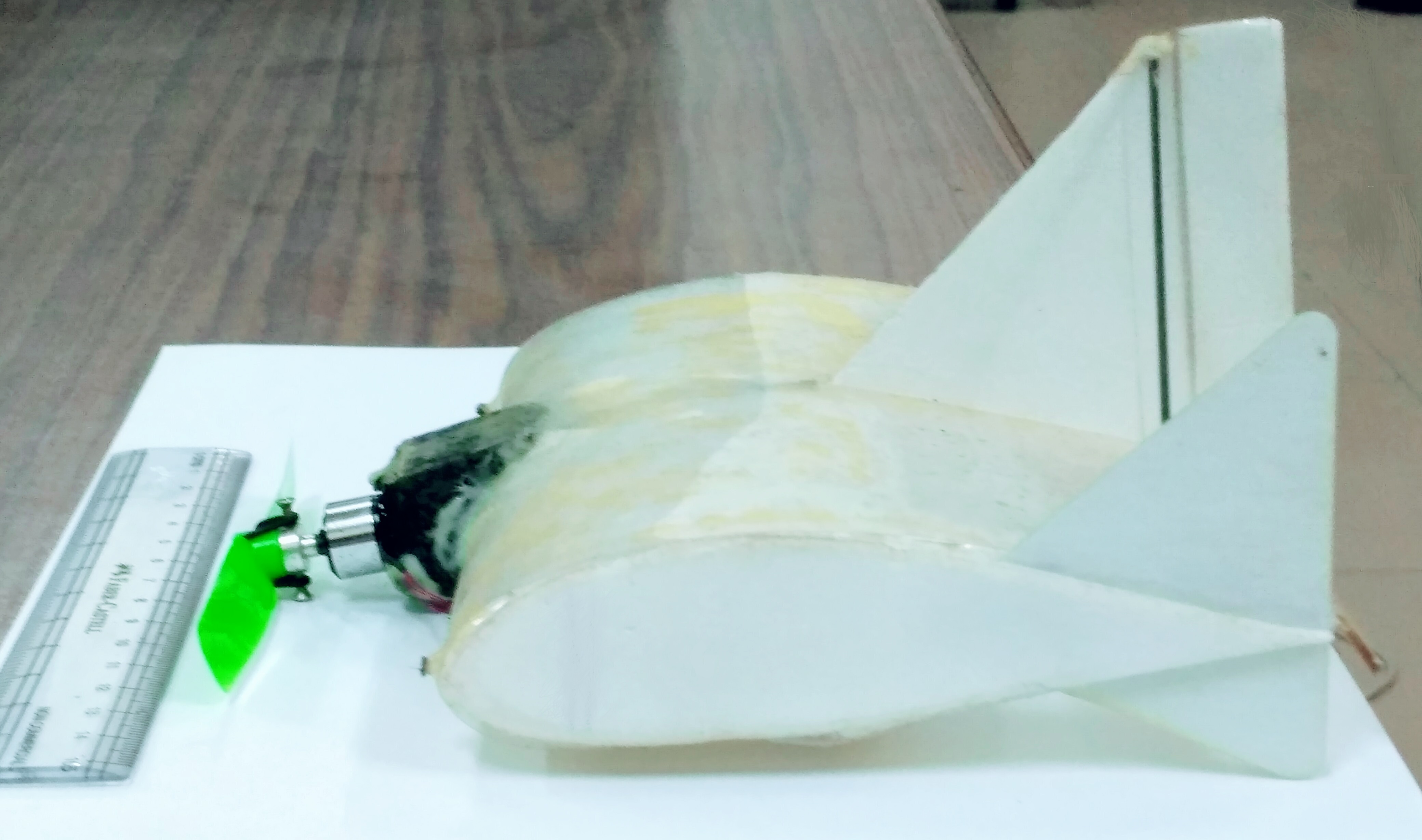}    
\caption{Photograph of KH2013A  MAV} 
\label{fig:mavfigure}
\end{flushright}
\end{figure}

 The lateral model of 150 mm KH2013A MAV is considered for simulation \cite{harikumar2016nonlinear}.  The photograph of KH2013A MAV is shown in Fig.\ref{fig:mavfigure}.
 In this MAV, the lateral control is done by rudder input($\delta_{r}$). The states, control and output variable are as follows:\\
$x_{p}=[v,p,r,\phi]';$ $\quad$
$u_{p}=\delta_{r} $;$\quad$
$y_{p}=\phi $.\\
\noindent The state matrix, control matrix, output matrix are as follows:\\ 
$ A_{lp}=\left[ \begin {array}{cccc} - 3.34& 1.93&- 7.55& 7.82
\\ \noalign{\medskip}- 40.5&- 2.22& 2.48& 0.0\\ \noalign{\medskip}
 234.0&- 2.84&- 27.0& 0.0\\ \noalign{\medskip} 0.0& 1.0& 0.268& 0.0
\end {array} \right] $ \\
$B_{lp}=[-8.41,59.7, 793.0, 0.0]'$; \quad
$C_{lp}=[0,0,0,1]$ \\

\noindent The available actuator bandwidth in case of MAV is limited due to restriction on weight and power budget. The actuator dynamics is considered as:
\begin{equation}
  \left[ \begin {array}{cccc} \dot{\delta}_{r}
\\ \noalign{\medskip} \ddot{\delta}_{r} \\ 
\end {array} \right] = \left[ \begin {array}{cccc} 0& 1
\\ \noalign{\medskip}-2367 & -72.22\\ \noalign{\medskip}
\end {array} \right]  \left[ \begin {array}{cccc} {\delta}_{r}
\\ \noalign{\medskip}\dot{\delta}_{r} \\ \noalign{\medskip}
\end {array} \right] + \left[ \begin {array}{cccc} 0
\\ \noalign{\medskip} 2367 \\ 
\end {array} \right] {\delta}_{command} \label{actuator}
\end{equation}
\noindent The augmented model of the plant  with actuator dynamics of (\ref{actuator}) is as follows:\\
States, control and output variable are 
$x_{p}=[v,p,r,\phi ,\delta_{r},\dot \delta_{r}]',$
$u_{p}=\delta_{command} $,
$y_{p}=\phi $.\\  
The augmented state, control and output matrix are\\
$ A_{p}=\left[ \begin {array}{cccccc} - 3.34& 1.93&- 7.55& 7.82 &-8.41 &0
\\ \noalign{\medskip}- 40.5&- 2.22& 2.48& 0.0 &59.7 &0\\ \noalign{\medskip}
 234.0&- 2.84&- 27.0& 0.0 &793.0 &0\\ \noalign{\medskip} 0.0& 1.0& 0.268& 0.0 & 0 &0 \\
 \noalign{\medskip} 0.0& 0& 0& 0.0 & 0 &1\\ \noalign{\medskip} 0.0& 0& 0 & 0.0 & -2367 &-72.22\\ 
\end {array} \right], $\\
\noindent $B_{p}=[0,0, 0, 0, 0,2367]'$; $\quad$
\noindent  $C_{p}=[0,0,0,1 ,0,0]$. 

 In this case, plant output  $\phi$ needs to track a single command  $ \phi_{cmd}$ using a single control input  $ \delta_{r} $. So, this tracking problem fits perfectly for SAC architecture.\\
 A first order BIBO stable simple reference model is considered. In case of SAC, the following reference model is considered.\\
  $\dot{x}_{m}=-5 x_{m}+5 \phi_{cmd}$; $\quad$ 
  ${y}_{m}= x_{m}$\\ 
\noindent In case of  CL-SAC, the reference model is as follows:\\

  $\dot{x}_{mo}=-5 x_{mo}+5 \phi_{cmd} -20(y_{mo}-y_{p} )$;  $\quad$
  ${y}_{mo}= x_{mo}$.\\  
  
\begin{figure}[H]
\begin{flushleft}
\includegraphics[width=1.0\linewidth, height=0.7\linewidth]{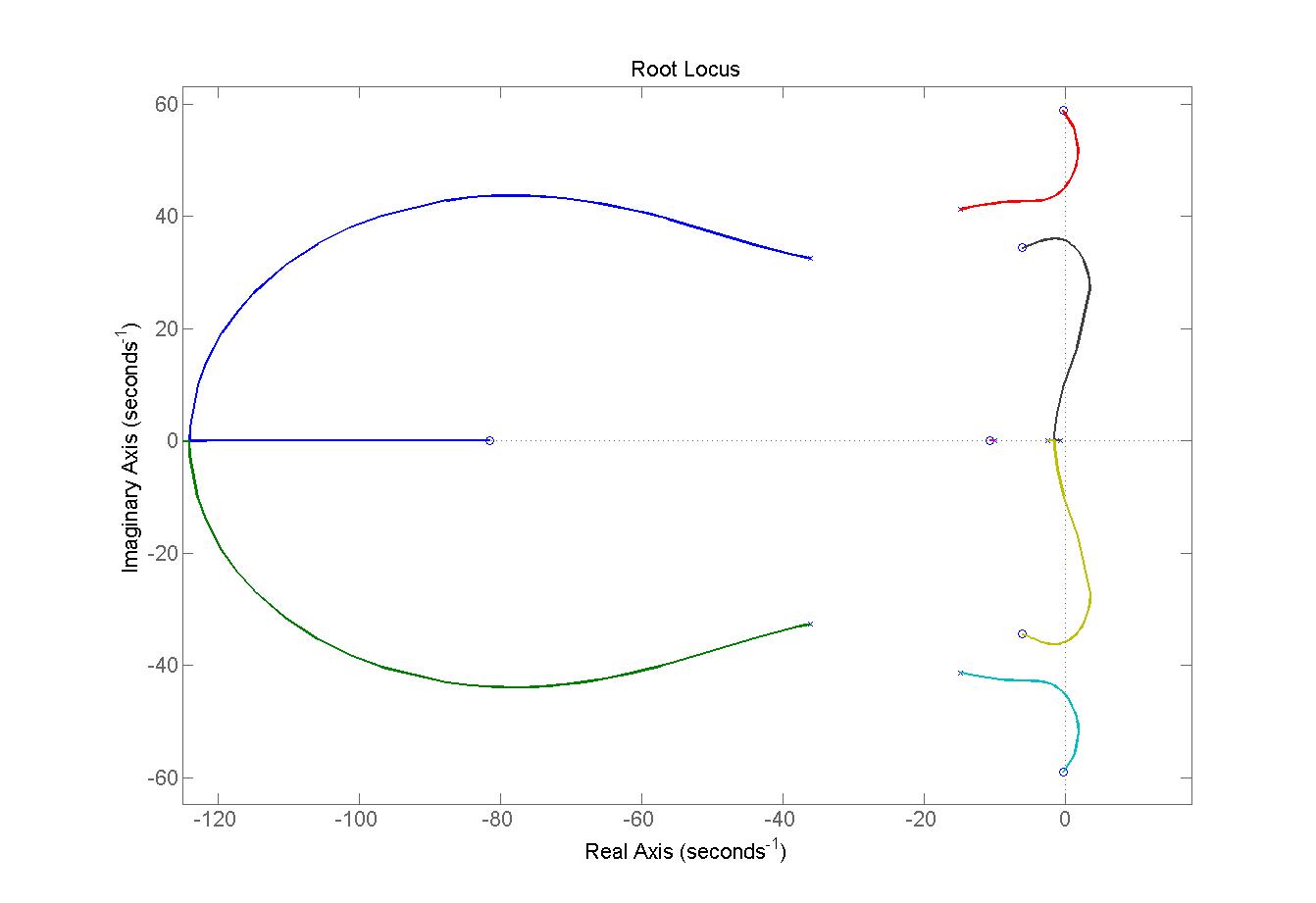}    
\caption{Root locus of plant with parallel feedforward } 
\label{fig:rlocus1}
\end{flushleft}
\end{figure}

\noindent So in this case,  
$A_{m}=[-5]$; $\quad$
$B_{m}=[5] $;$\quad$
$C_{m}=[1]$;$\quad$
$u_{m}=\phi_{cmd}$; $\quad $ $L_{v}$=20 are considered. Also, the dimension of the considered reference model is smaller than the plant model and the reference model output is the same as number of  output variable to be tracked.
The augmented plant is not W-ASPR. So, parallel feedforward configuration is used to make the plant W-ASPR. In this case the plant transfer function $T(s)$= $C_{p}(sI-A_{p})^{-1} B_{p}$:  $\frac{\text{num}(s)}{\text{den}(s)}$; where,\\
\begin{equation*}
\text{num}(s)= 6.444e05 s^2 + 1.119e07 s + 9.185e08
\end{equation*}
\begin{eqnarray*}
\text{den}(s)= s^6 + 104.8 s^5 + 6728 s^4 + 2.28e05 s^3  + 5.18e06 s^2 \\+ 1.4e07 s + 6.351e06. 
\end{eqnarray*}
\noindent $C(s)$ is  chosen such that it stabilized the plant $T(s)$ and its gain is selected as high as possible while ensuring the stability.
\noindent  In this case, after root locus analysis $C(s)$ is chosen as : $\frac{4s+40}{10}$ .
\noindent  The parallel feedforward transfer functions is selected as:
$ D(s) =\frac{1}{C(s)}$.
 \noindent  The plant with feedforward transfer function $T(s) + D(s)$ is:
 $F(s)=\frac{\text{num}_F(s)}{\text{den}_F(s)}$; where, 
 \begin{eqnarray*}
 \text{num}_F(s)= 10 s^6 + 1048 s^5 + 6.7e04 s^4 + 4.8e06 s^3 \\ + 1.2e08 s^2 + 4.2e09 s + 3.68e10
 \end{eqnarray*}
 \begin{eqnarray*}
 \text{den}_F(s)= 4 s^7 + 459.1 s^6 + 3.11e04 s^5 + 1.1e06 s^4  \\+ 2.9e07 s^3 + 2.6e08 s^2 + 5.8e08 s  + 2.5e08 
 \end{eqnarray*}
 \noindent The root locus of $T(s) + D(s)$ is shown in Figure \ref{fig:rlocus1}. Clearly, the augmented plant is W-ASPR.

\begin{figure}[H]
\begin{flushright}
\includegraphics[width=1.1\linewidth, height= 0.7\linewidth]{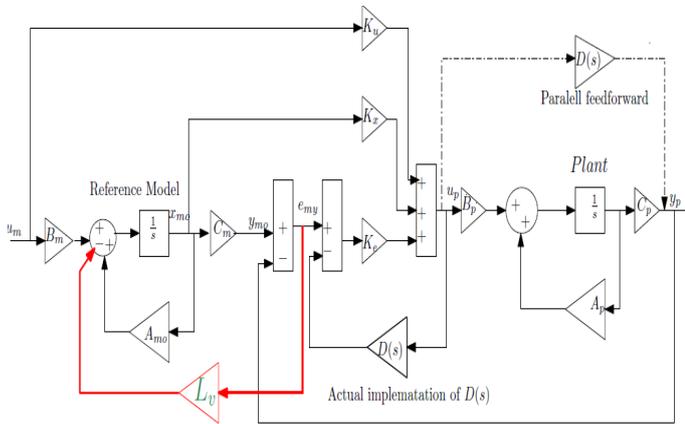}    
\caption{Details of CL-SAC architecture} 
\label{fig:clsacimg}
\end{flushright}
\end{figure}

 \noindent  The details of implementation in case of CL-SAC are shown in Fig. \ref{fig:clsacimg}.                                                                     The parallel feedforward compensator $D(s)$ cannot be implemented parallel to the plant ( shown in dotted line in  Fig. \ref{fig:clsacimg} ); therefore,  the equivalent representation of $D(s)$ is considered through the solid line path.
 
The value of $\sigma $ is  chosen as 5 . The value of  adaptation gains $ \Gamma_{pe}$ , $\Gamma_{Ie}$ , $\Gamma_{px}$ , $\Gamma_{Ix}$ , $\Gamma_{pu}$ , $\Gamma_{Iu}$ are chosen as 10 in simulation .\\

\begin{figure}[H]
\begin{flushright}
\includegraphics[width=1.0\linewidth, height= 1.0\linewidth]{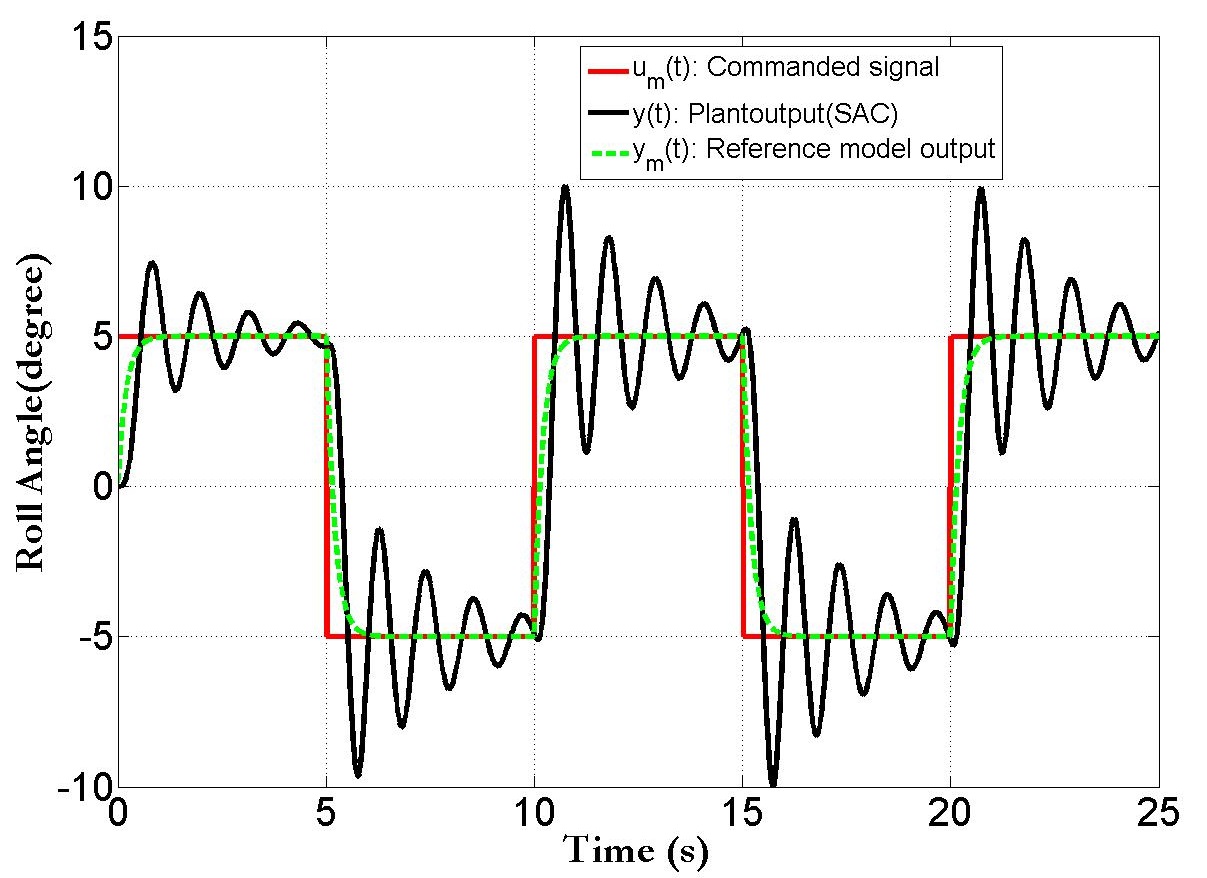}    
\caption{Reference signal tracking using SAC architecture } 
\label{fig:fig3}
\end{flushright}
\end{figure}

\begin{figure}
\begin{flushright}
\includegraphics[width=1.0\linewidth, height= 1.0\linewidth]{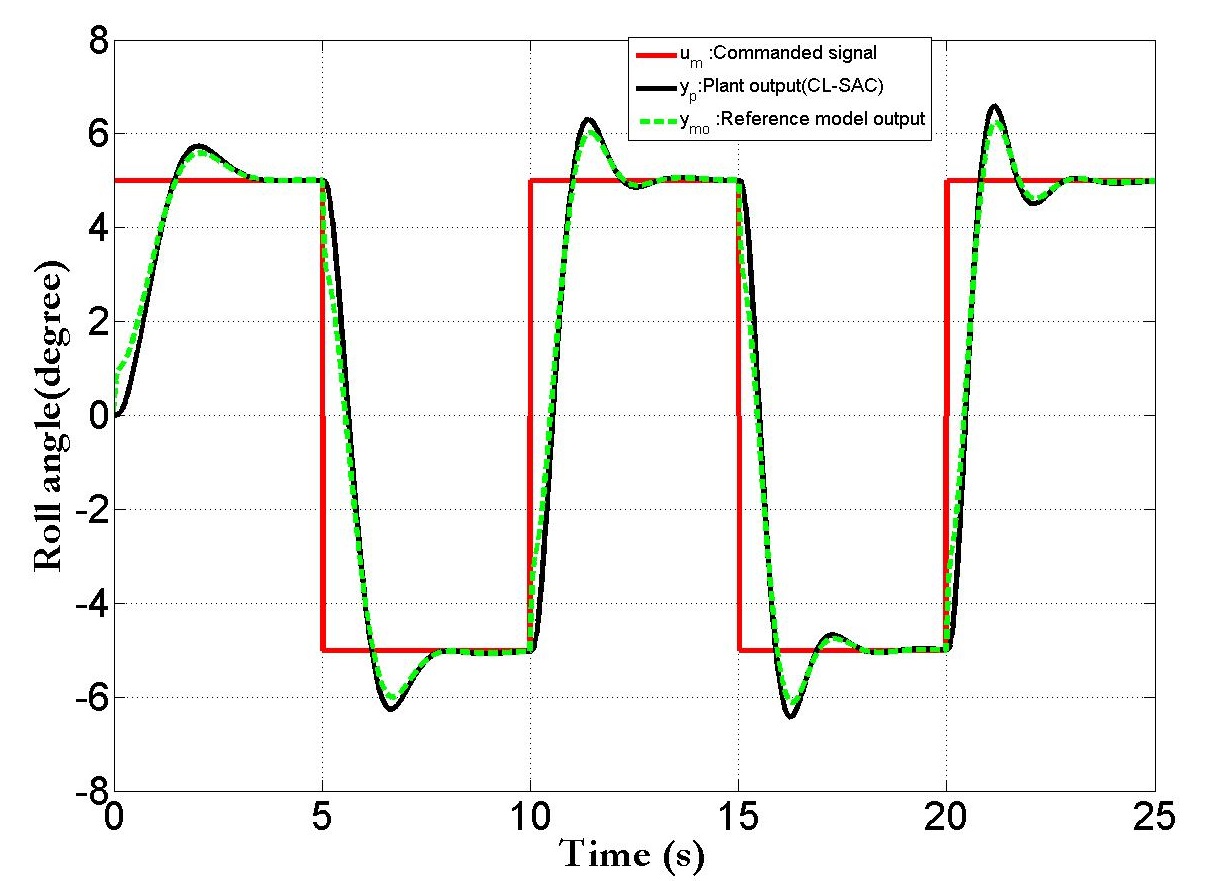}    
\caption{Reference signal tracking using CL-SAC architecture} 
\label{fig:fig4}
\end{flushright}
\end{figure}

The tracking of the commanded square wave by the SAC and the CL-SAC is plotted in Fig.\ref{fig:fig3} and Fig.\ref{fig:fig4} respectively.

\begin{figure}[H]
\begin{flushright}
\includegraphics[width=1.0\linewidth, height= 1.0\linewidth]{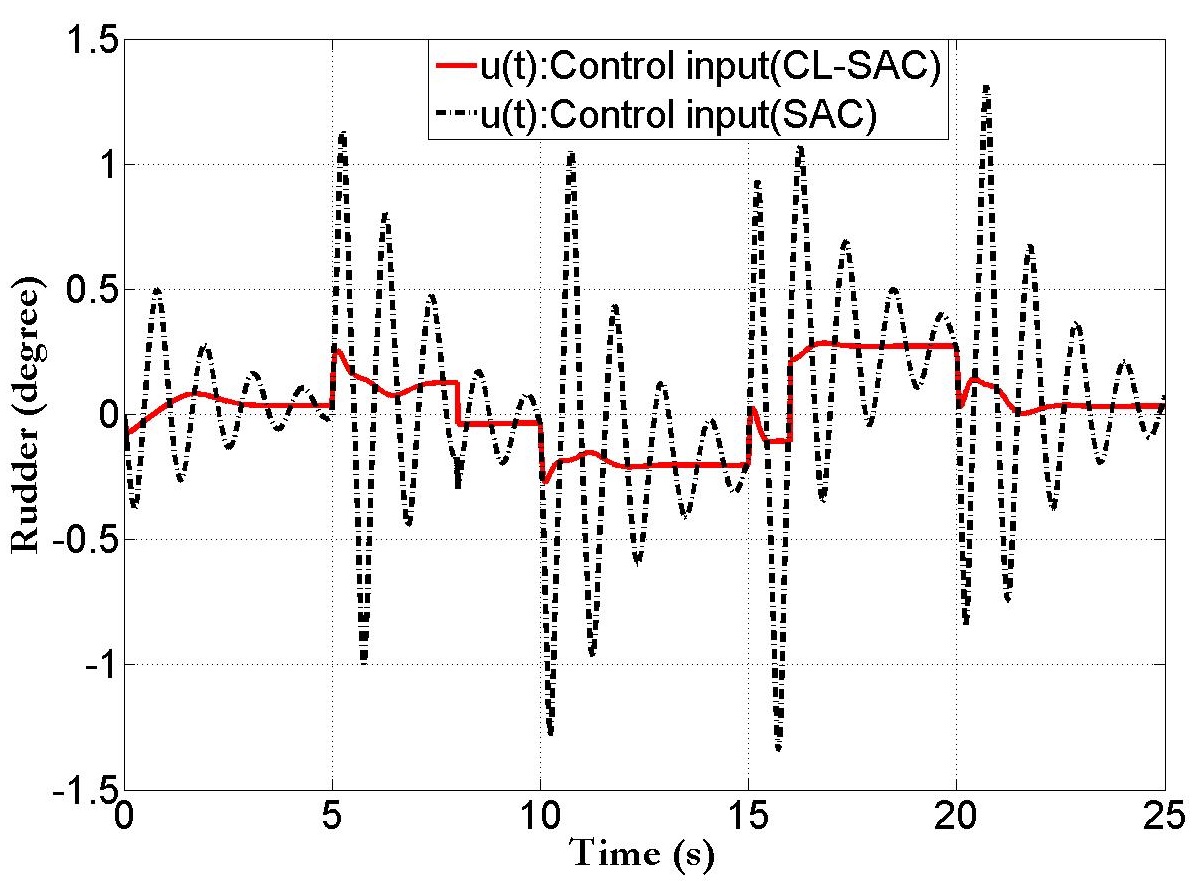}    
\caption{Figure comparing the control input $(u_{p})$ between SAC and CL-SAC} 
\label{fig:fig5}
\end{flushright}
\end{figure}

\begin{figure}[H]
\begin{flushright}
\includegraphics[width=1.0\linewidth, height= 1.0\linewidth]{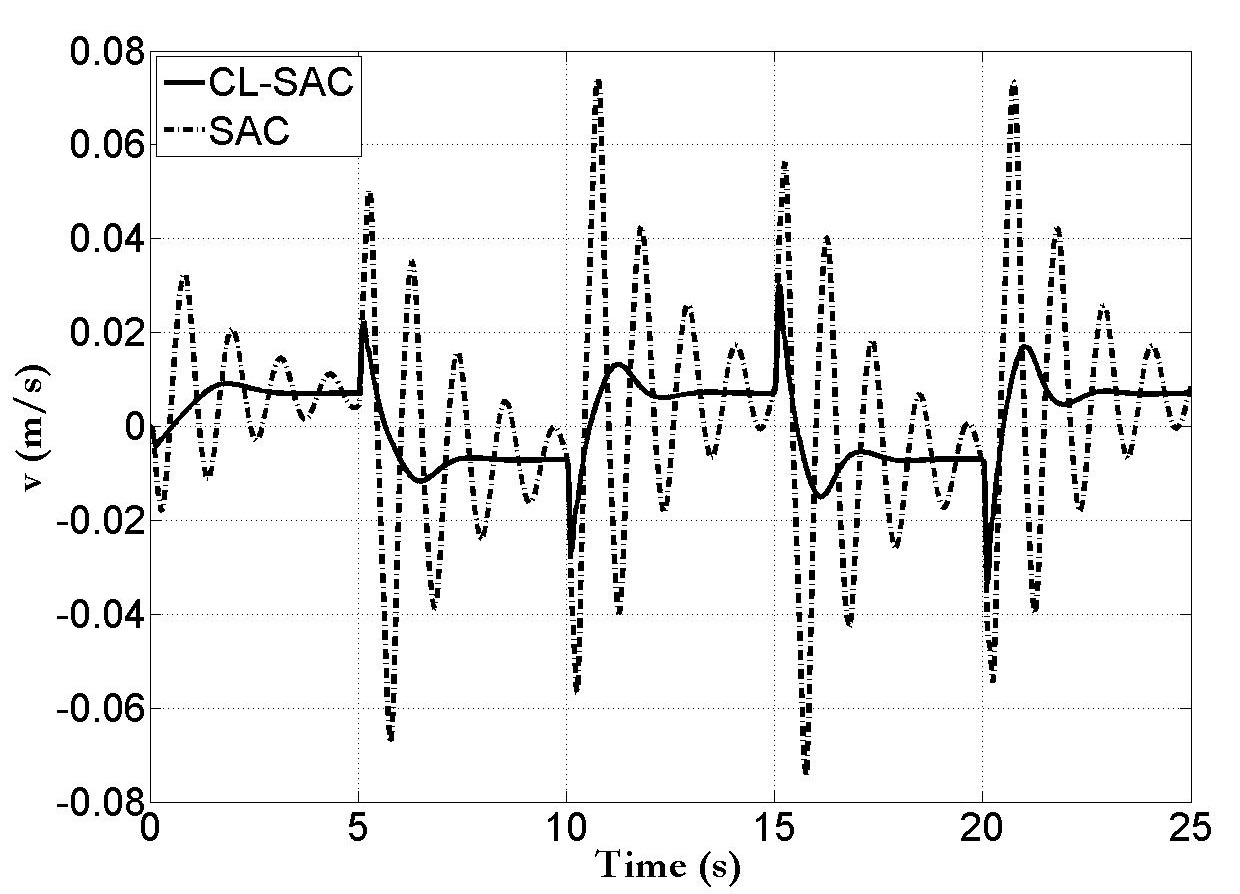}    
\caption{Figure comparing the lateral velocity between SAC and CL-SAC} 
\label{fig:fig6}
\end{flushright}
\end{figure}

\begin{figure}[H]
\begin{flushright}
\includegraphics[width=1.0\linewidth, height= 1.0\linewidth]{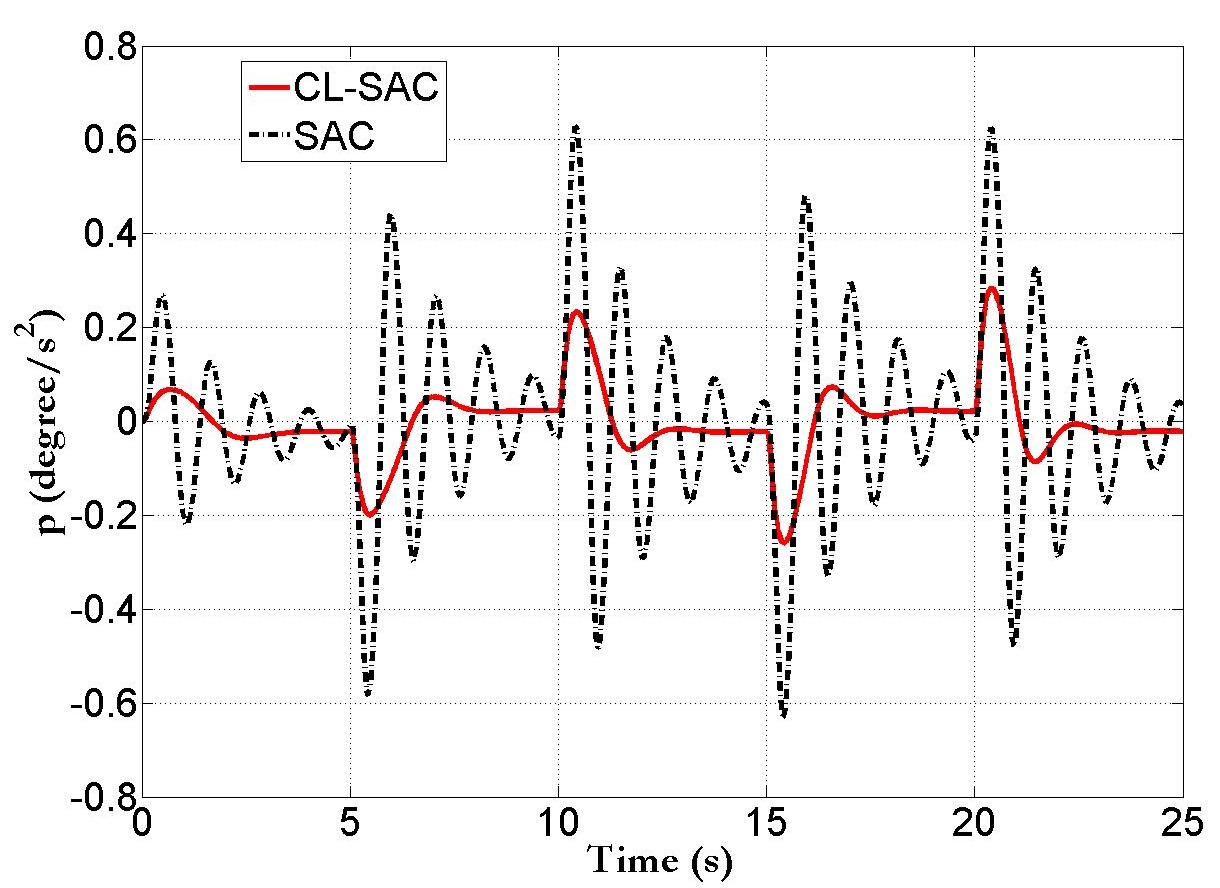}    
\caption{Figure comparing the roll rate between SAC and CL-SAC} 
\label{fig:fig7}
\end{flushright}
\end{figure}

\begin{figure}[H]
\begin{flushright}
\includegraphics[width=1.0\linewidth, height= 1.0\linewidth]{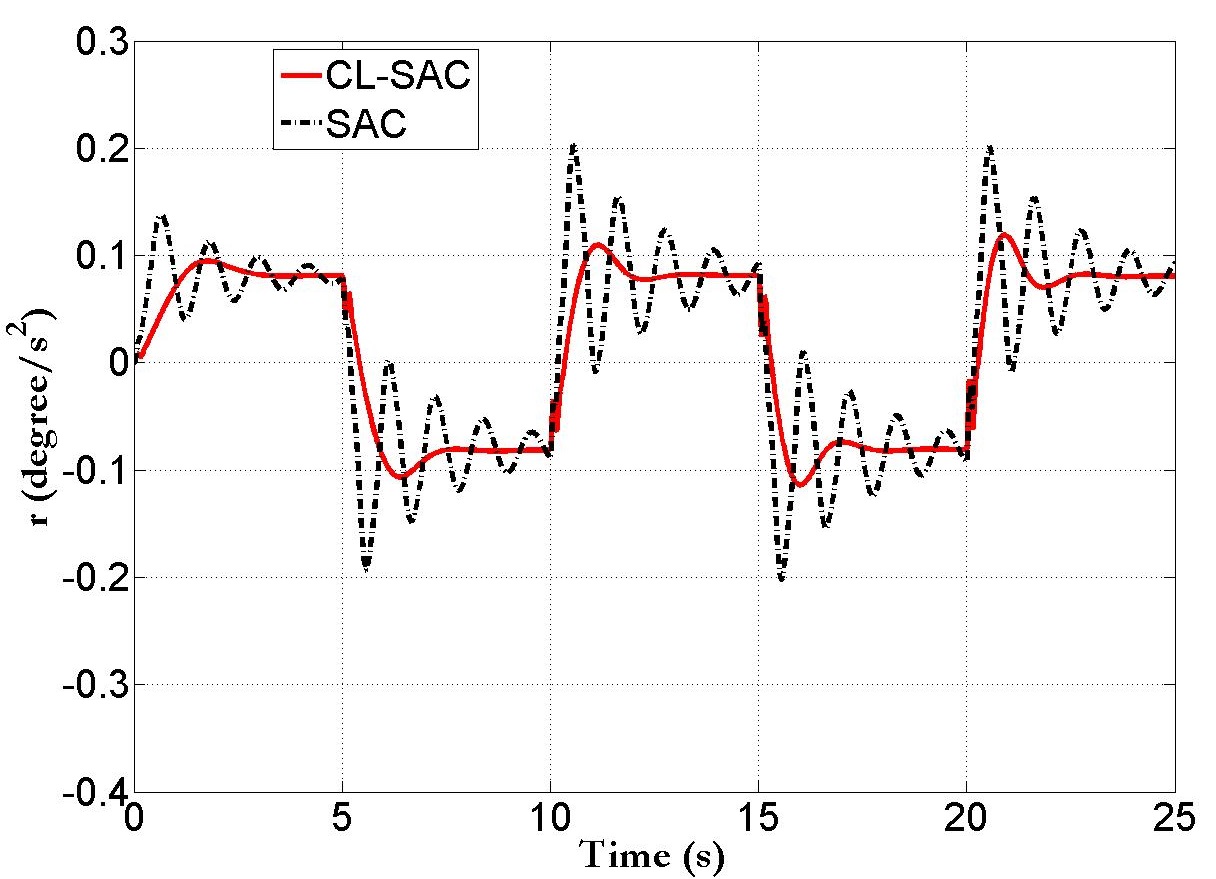}    
\caption{Figure comparing the yaw rate between SAC and CL-SAC} 
\label{fig:fig8}
\end{flushright}
\end{figure}

\begin{figure}[H]
\begin{flushright}
\includegraphics[width=1.0\linewidth, height= 1.0\linewidth]{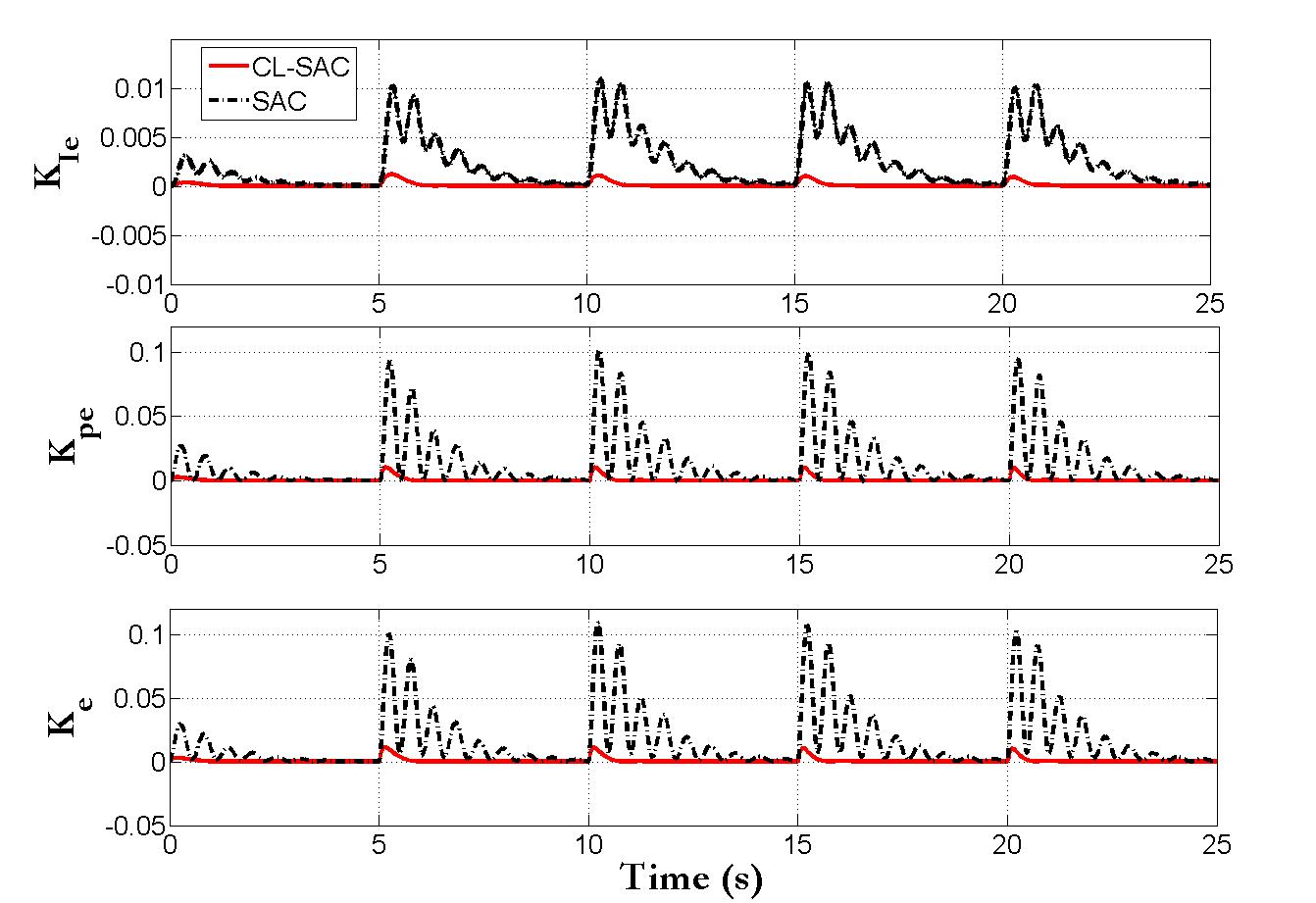}    
\caption{Figure comparing the value of $K_{Ie}$, $K_{pe}$, $K_{e}$ between SAC and CL-SAC} 
\label{fig:ke}
\end{flushright}
\end{figure}

\begin{figure}[H]
\begin{flushright}
\includegraphics[width=1.0\linewidth, height= 1.0\linewidth]{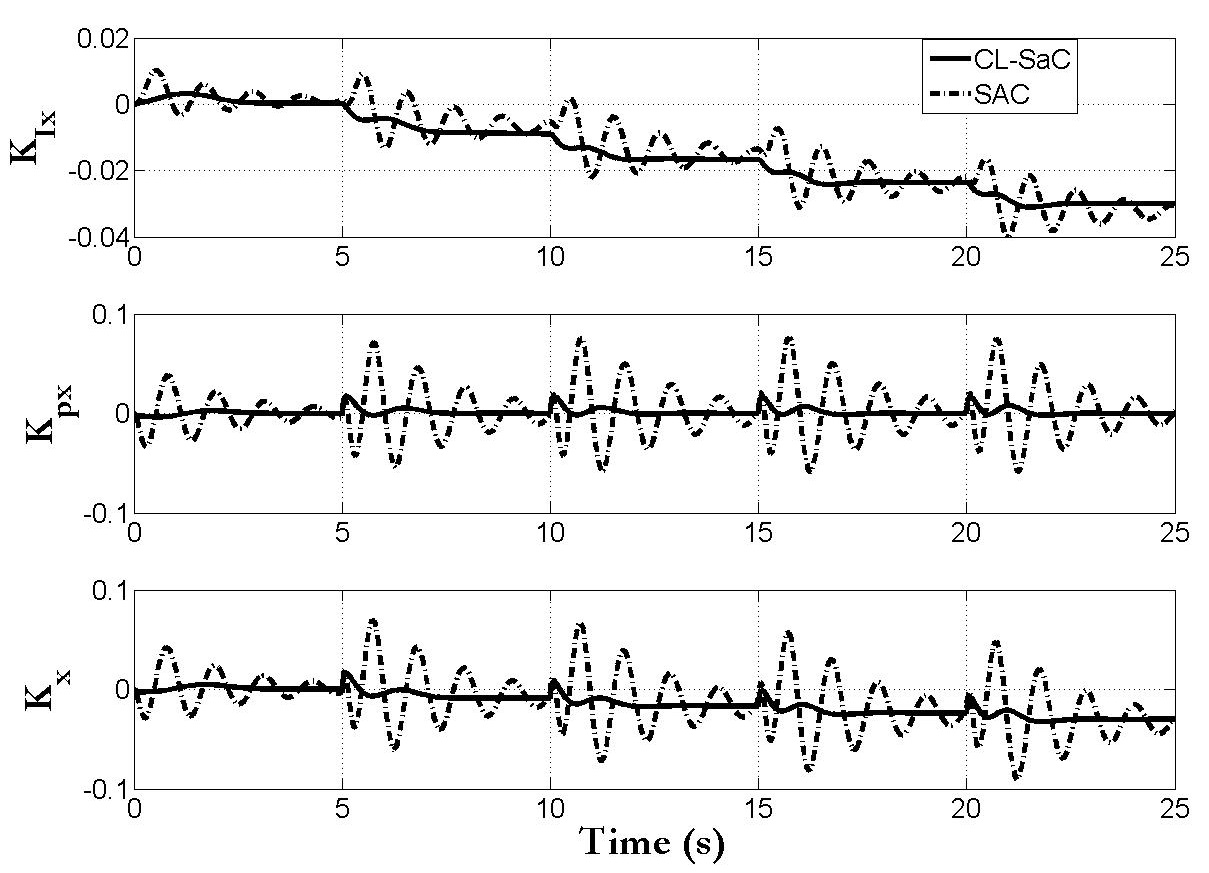}    
\caption{ Figure comparing the value of $K_{Ix}$, $K_{px}$, $K_{x}$ between SAC and CL-SAC} 
\label{fig:kx}
\end{flushright}
\end{figure}

\begin{figure}[H]
\begin{flushright}
\includegraphics[width=1.0\linewidth, height= 1.0\linewidth]{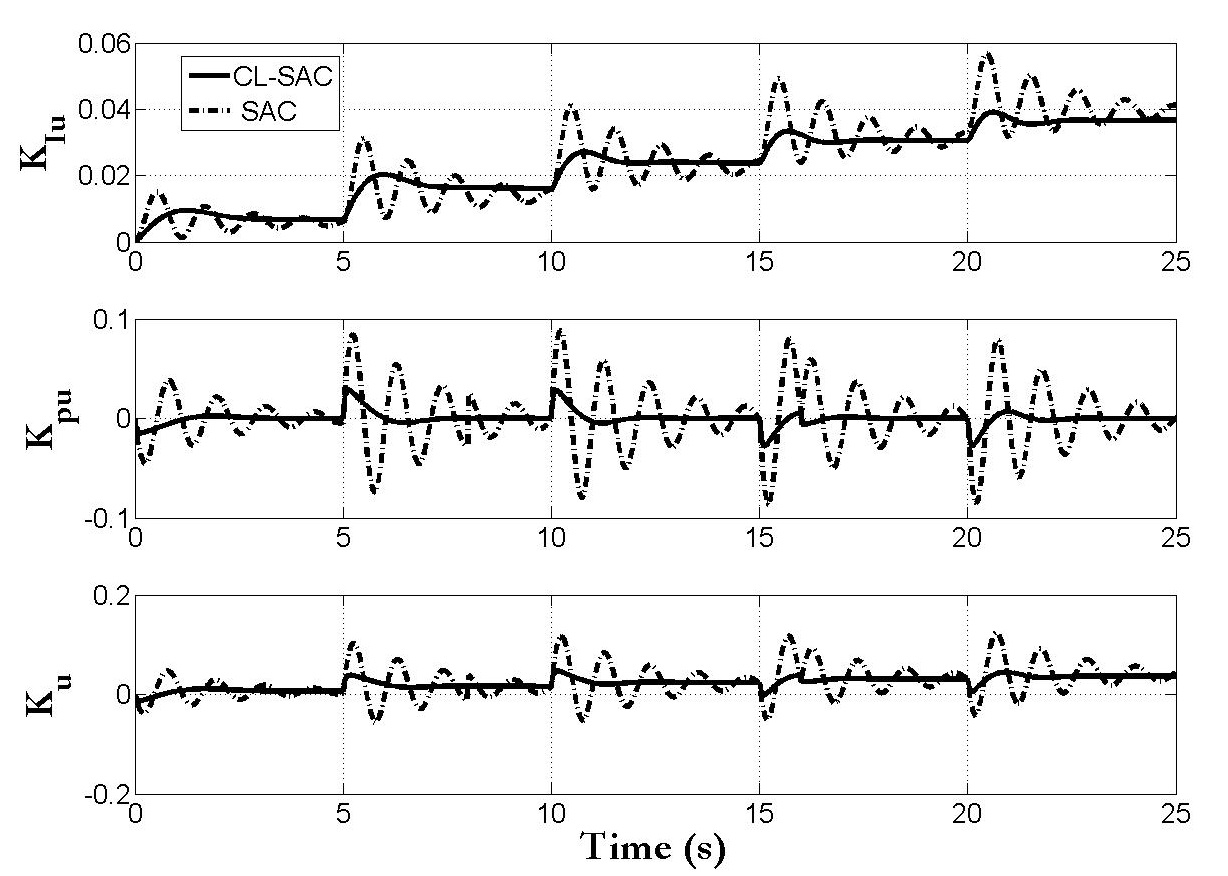}    
\caption{Figure comparing value of $K_{Iu}$, $K_{pu}$, $K_{u}$ between SAC and CL-SAC} 
\label{fig:ku}
\end{flushright}
\end{figure}

\begin{figure}
\begin{flushright}
\includegraphics[width=1.2\linewidth, height=1.0 \linewidth]{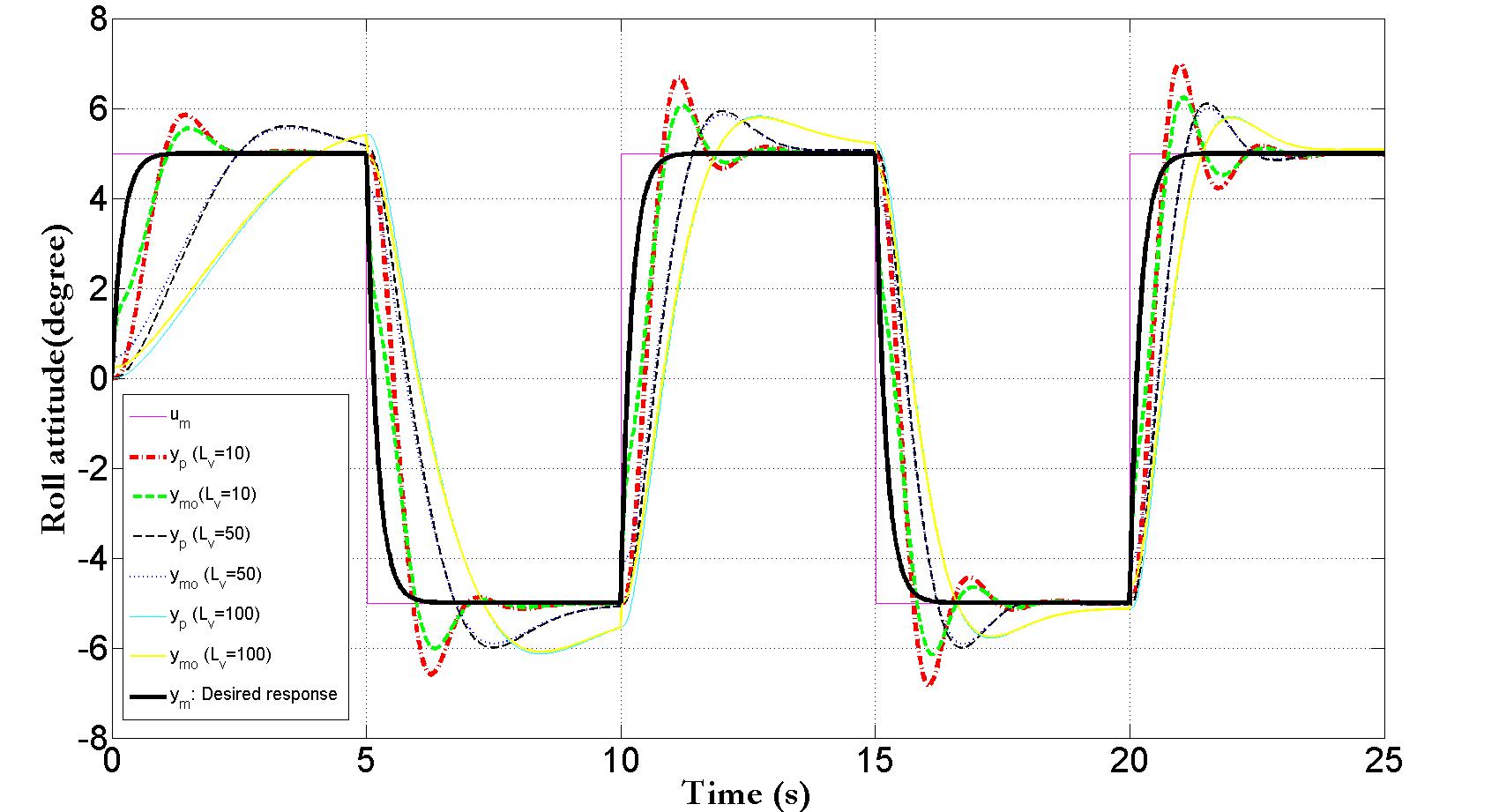}    
\caption{Figure comparing the tracking response for different value of $L_{v}$} 
\label{fig:lvvartracking}
\end{flushright}
\end{figure}

\begin{figure}
\begin{flushright}
\includegraphics[width=1.2\linewidth, height= 1.0\linewidth]{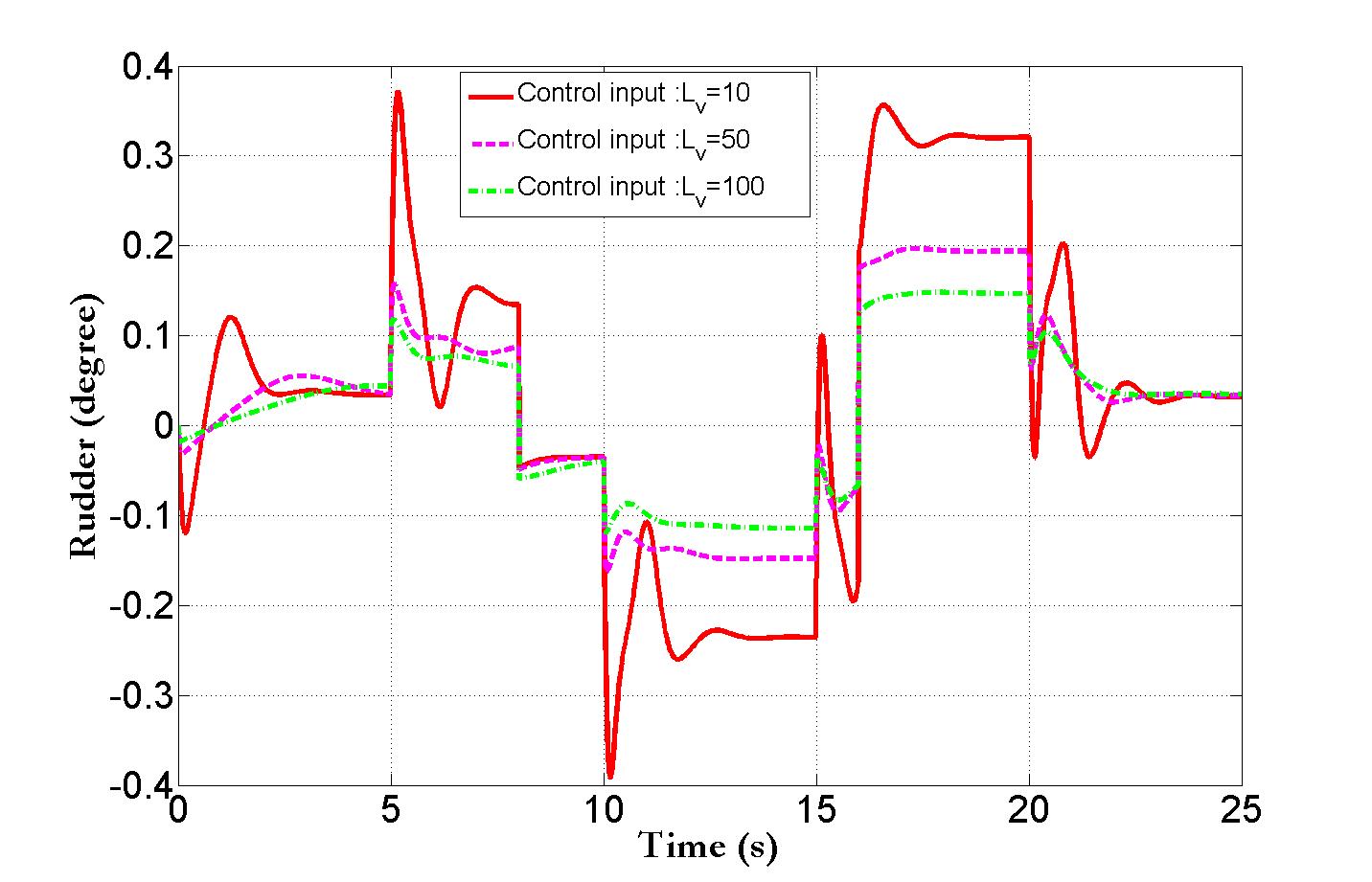}    
\caption{Figure comparing control response for different value of $L_{v}$} 
\label{fig:lvvarcontrol}
\end{flushright}
\end{figure}

From Fig.\ref{fig:fig3} and Fig.\ref{fig:fig4}, it is clear that the tracking performance improves in case of CL-SAC as compared to SAC. Fig. \ref{fig:fig5} compares the control efforts between both architecture. 
  The control efforts in CL-SAC is also less as compared to that of SAC in  Fig.\ref{fig:fig5}. Clearly, improvement of the transient performance in CL-SAC is obtained without extra control efforts. 
  The variation of other  states lateral velocity (v), roll rate (p) and  yaw rate (r) are plotted in Fig.\ref{fig:fig6},  Fig.\ref{fig:fig7} and Fig.\ref{fig:fig8}.
From Fig.\ref{fig:fig6},  Fig. \ref{fig:fig7} and Fig.\ref{fig:fig8}; it is evident that the variation of other states  in CL-SAC structure is smoother than SAC structure.
 The control gains $ K_{e}$, $ K_{x}$, $K_{u}$ are plotted in  Fig.\ref{fig:ke}, Fig.\ref{fig:kx} and  Fig.\ref{fig:ku}.
The value of $L_{v} $ needs to be chosen appropriately. The variation in tracking response is investigated with three different values of output error feedback gain; $L_{v}$  =10, 50 and 100.
The variation in tracking performance and control input with the variation of ($L_{v}$) is shown in Fig.\ref{fig:lvvartracking} and  Fig.\ref{fig:lvvarcontrol}.
Clearly from Fig.\ref{fig:lvvartracking}, the error dynamics between the plant output and the close loop reference model output becomes faster with the increase in the value of ($L_{v}$), however, it causes poor tracking of the output of the open loop reference model; hence poor tracking of original reference command. The lower value of $L_{v} $ will have less effect on error feedback term, where the high value of $L_{v} $ can cause poor tracking of the output of open loop reference model. As output error reduces with a higher value of ($L_{v}$), the control effort also reduces; which is evident from Fig.\ref{fig:lvvarcontrol}. 
\subsection{Discussions}
From the simulation of lateral model dynamics, it is clear that the roll attitude tracking performance during transient phase is better in the case of CL-SAC as compared to SAC without additional control efforts. So, the output error feedback term in reference model dynamics improves the performance of the system. However, the gain of the output error feedback term needs to be chosen properly as per performance specification and available actuator.  The optimum design of $L_{v}$ for CL-SAC structure should be investigated in future studies. 
\section{Conclusions}
In this paper new CL-SAC architecture is proposed after modification of  SAC architecture with closed loop reference model based on output feedback. The proposed architecture is based on the intuition that output tracking error feedback in the reference model dynamics will  help the plant to follow the reference model.
It is shown through mathematical analysis and simulations of lateral model dynamics of Micro Air Vehicle that CL-SAC improves the transient performance without additional control requirement.  

\begin{acknowledgements}
The authors would like to thank ARDB and NPMICAV program for partial funding of this work.
\end{acknowledgements}

\bibliographystyle{spmpsci}      

\bibliography{template}

\end{document}